\documentclass{aastex}
\usepackage{spr-astr-addons}
\usepackage{url}
\usepackage{epstopdf}
\usepackage{txfonts}
\usepackage{graphicx}

\RequirePackage{color}

\begin{document}

\title{On the nature of dark energy: the lattice Universe}
\slugcomment{Not to appear in Nonlearned J., 45.}
\shorttitle{On the nature of dark energy: the lattice Universe}
\shortauthors{M.~Villata}

\author{M.~Villata}
\affil{INAF, Osservatorio Astrofisico di Torino, Via Osservatorio 20, I-10025 Pino Torinese (TO), Italy\\
e-mail: villata@oato.inaf.it}

\begin{abstract}
There is something unknown in the cosmos. Something big. Which causes the acceleration of the Universe expansion, that is perhaps the most surprising and unexpected discovery of the last decades, and thus represents one of the most pressing mysteries of the Universe. The current standard $\Lambda$CDM model uses two unknown entities to make everything fit: dark energy and dark matter, which together would constitute more than 95\% of the energy density of the Universe. A bit like saying that we have understood almost nothing, but without openly admitting it. Here we start from the recent theoretical results that come from the extension of general relativity to antimatter, through CPT symmetry. This theory predicts a mutual gravitational repulsion between matter and antimatter. Our basic assumption is that the Universe contains equal amounts of matter and antimatter, with antimatter possibly located in cosmic voids, as discussed in previous works. From this scenario we develop a simple cosmological model, from whose equations we derive the first results. While the existence of the elusive dark energy is completely replaced by gravitational repulsion, the presence of dark matter is not excluded, but not strictly required, as most of the related phenomena can also be ascribed to repulsive-gravity effects. With a matter energy density ranging from $\sim5\%$ (baryonic matter alone, and as much antimatter) to $\sim25\%$ of the so-called critical density, the present age of the Universe varies between about 13 and $15\rm\,Gyr$. The SN Ia test is successfully passed, with residuals comparable with those of the $\Lambda$CDM model in the observed redshift range, but with a clear prediction for fainter SNe at higher $z$. Moreover, this model has neither horizon nor coincidence problems, and no initial singularity is requested. In conclusion, we have replaced all the tough problems of the current standard cosmology (including the matter-antimatter asymmetry) with only one question: is the gravitational interaction between matter and antimatter really repulsive as predicted by the theory and as the observation of the Universe seems to suggest? We are awaiting experimental responses.
\end{abstract}



\section{Introduction}

Since the discovery of the cosmic expansion acceleration in 1998 \citep[e.g.][]{rie98,per99}, one of the most debated questions in physics and cosmology has been the existence and nature of the so-called dark energy, which should account for that unexpected phenomenon. Indeed, a repulsive force acting in the Universe space-time defies any previous physical knowledge, as the only known interaction among matter on these large scales is the universal Newton-Einstein gravitational attraction. Both the classical (Newtonian) and relativistic (Einsteinian) theories of gravitation seem to exclude that gravity can be in some way repulsive. However, in two recent papers \citet{vil11,vil12b} showed that the general theory of relativity can be consistently extended to the existence of antimatter (which was unknown at the epoch of the birth of the two theories), based on its CPT properties, which imply that matter and antimatter are both gravitationally self-attractive, but mutually repulsive. Thus, if our Universe contains a certain amount of antimatter (possibly equivalent to the matter one, due to the expected matter-antimatter symmetry), the origin of the cosmic speed-up can be easily and naturally explained (together with the well-known expansion itself), without any need of mysterious ingredients like a physically unknown dark energy, or of modifications to the current well-established theories.

Thus far, the Universe acceleration, i.e.\ the presence of the elusive dark energy, has been formally ascribed to an additional term having a negative pressure in the cosmic-expansion equations, in the simplest case corresponding to a cosmological constant, perhaps associated to the energy of the quantum vacuum. Besides this standard cosmology of the $\Lambda$CDM model, various alternatives have been proposed to explain the accelerated expansion, invoking scalar fields or modifications of general relativity, such as extensions to extra dimensions or higher-order curvature terms \citep[e.g.][]{ame00,dva00,car04,cap05,nap12}. For an extensive and detailed discussion (and bibliography) on alternative models of dark-energy cosmologies see the recent review by \citet{bam12}.

The main problem with the connection of dark energy with the quantum vacuum energy is that the latter would be expected to be some $10^{120}$ times larger than observed, or at least $\sim10^{40}$ when considering only quantum chromodynamics \citep[e.g.][]{wei89}. Recently, starting from the assumption of repulsive gravity between matter and antimatter, \citet{haj12b} has shown that dark energy could be the result of the gravitational polarization of the quantum vacuum, while in previous papers \citep{haj11c,haj12a} the author showed that also dark matter could be an illusion caused by the same phenomenon. This treatment of quantum vacuum virtual pairs leads to much more reasonable theoretical values for the associated dark energy, but not yet in agreement with observational estimates. Another big problem in this sense is the so-called ``coincidence" problem \citep[e.g.][and references therein]{pee03}: why should the dark energy density be so comparable (a small factor larger) with the energy density of matter, and why just now in the history of the Universe (while, being constant in the $\Lambda$CDM model, in the past it would have been so negligible and in the future it will dominate)? Thus, one suspects that actually the repulsive force could be closely related to the matter content, as we will show in this paper. In Sect.\ 2 we present the basis of general relativity extended to the presence of antimatter, while in Sect.\ 3 we apply it in a cosmological model, whose results are submitted to the main observational tests in Sect.\ 4. Conclusions are drawn in Sect.\ 5.

\section{Antimatter gravity}

Antimatter gravity has a long and troubled history, which begins with the belief that gravity can be only attractive and never repulsive, then this concept passes through phases of questioning and contestation, to reach a growing consensus towards the possibility that the gravitational interaction between matter and antimatter is repulsive. The main steps of this story can be drawn from many works of various authors, e.g.\ \citet{mor58,sch58,sch59,goo61,nie91,noy91,cha92,cha93,cha97,ni04,noy08,cab10,cab11,haj10c,haj11c,haj11d,haj12a,haj12b,vil11,vil12a,vil12b,ben12,dop12}; and references therein. 

Here we start from the assumption that antimatter is CPT-transformed matter and, as already partially done in \citet{vil11,vil12b}, show how gravitational repulsion between matter and antimatter is a natural outcome of general relativity, and derive the relevant field and cosmological equations.

\subsection{Equation of motion}

In a metric theory of space-time, the action for a free test particle with rest mass $m$ and line element ${\rm d}s$ is defined as
\begin{equation}
\label{eq.1}
S=-m\int{{\rm d}s}=-m\int{\sqrt{g_{\mu\nu}{\rm d}x^\mu{\rm d}x^\nu}}\,.
\end{equation}
With $x^0=t$ and $\dot{x}^\mu={\rm d}x^\mu/{\rm d}t$, in terms of the Lagrangian $L$ we have
\begin{equation}
\label{eq.2}
S=\int{L\,{\rm d}t}\,,\quad L=-m\frac{{\rm d}s}{{\rm d}t}=-m\sqrt{g_{\mu\nu}\dot{x}^\mu\dot{x}^\nu}\,.
\end{equation}
With this classical definition of the Lagrangian, the line element (or the action) of the test particle, which is a scalar, is split into two non-scalar parts, ${\rm d}s/{\rm d}t$ (or $L$) and ${\rm d}t$, which in particular are both (CP)T-odd, i.e.\ their CPT-transformed counterparts change sign.

Under the assumption that antimatter is CPT-transformed matter \citep{vil11}, while for matter (M) ${\rm d}s/{\rm d}t$ and ${\rm d}t$ are both positive definite, for antimatter (A) they are both negative definite. Consequently, we have
\begin{equation}
\label{eq.3}
L_{\rm A}=-m\frac{{\rm d}s}{-{\rm d}t}=+m\sqrt{g_{\mu\nu}\dot{x}^\mu\dot{x}^\nu}=-L_{\rm M}\,.
\end{equation}

The canonical momenta are defined as 
\begin{equation}
\label{eq.4}
p_i=\frac{\partial L}{\partial\dot{x}^i}=-mg_{i\mu}u^\mu\,,
\end{equation}
where $u^\mu={\rm d}x^\mu/{\rm d}s$. Apart from the sign, they are the three spatial components of the covariant version of the energy-momentum four-vector. The time component is given by the Hamiltonian
\begin{equation}
\label{eq.5}
H=p_i\dot{x}^i-L=-mg_{i\mu}u^\mu\dot{x}^i+m\frac{{\rm d}s}{{\rm d}t}=mg_{0\mu}u^\mu=-p_0\,.
\end{equation}
All the terms in Eqs.~(\ref{eq.4}) and (\ref{eq.5}) are (C)PT-odd, as expected, since four-vectors are always (C)PT-odd, and thus change sign from matter to antimatter.

Through the Euler-Lagrange equation of motion and the energy equation,
\begin{equation}
\label{eq.6}
\frac{{\rm d}p_i}{{\rm d}t}=\frac{\partial L}{\partial x^i}\,,\quad\frac{{\rm d}H}{{\rm d}t}=\frac{\partial H}{\partial t}\,,
\end{equation}
which are both (C)PT-invariant with all terms (C)PT-even, we can get the well-known geodesic equation, i.e.\ the four-component equation of motion of general relativity,
\begin{equation}
\label{eq.7}
\frac{{\rm d}u^\lambda}{{\rm d}s}=-\Gamma^\lambda_{\mu\nu}u^\mu u^\nu\,,
\end{equation}
where the Christoffel symbol $\Gamma^\lambda_{\mu\nu}$ represents the (matter-generated) gravitational field.

As pointed out by \citet{vil11}, this equation is composed of four (C)PT-odd elements. If we CPT-transform all the four elements, we obtain an identical equation describing the motion of an antimatter test particle in an antimatter-generated gravitational field, since all the four changes of sign cancel one another. Thus, this CPT symmetry ensures the same self-attractive gravitational behavior for both matter and antimatter. However, if we transform only one of the two components, either the field $\Gamma^{\lambda}_{\mu\nu}$ or the particle (represented by the remaining three elements), we get a change of sign that converts the original gravitational attraction into repulsion, so that matter and antimatter repel each other\footnote{The geodesic equation for a massless particle, such as a photon, is formally equal to Eq.~(\ref{eq.7}), except for the parameter $s$, which can no longer be taken as the proper time, being ${\rm d}s=0$, but it will be an affine parameter describing the world line. Thus, a (retarded) photon will be repelled by an antimatter-generated gravitational field, and a CPT-transformed photon, i.e.\ an advanced photon, will be repelled by matter.}.

\subsection{Field equations}

In order to obtain a generally covariant field equation, one must derive it from an action composed of a scalar Lagrangian density $\cal L$ multiplied by the scalar factor $\sqrt{-g}\,{\rm d}^4x$, which guarantee the scalarness of the action itself:
\begin{equation}
\label{eq.8}
S=\int{\sqrt{-g}\,{\cal L}\,{\rm d}^4x}\,.
\end{equation}

To get the action for a point particle in this form, one usually multiplies the time integral in Eq.~(\ref{eq.2}) by a space integral over the three-dimensional Dirac delta function $\delta^3({\bf x}-{\bf x}(t))$, where ${\bf x}(t)$ is the position of the particle at time $t$. Given a system of these particles, its action will be
\begin{equation}
\label{eq.9}
S=-\sum_n{m_n\int{\delta^3({\bf x}-{\bf x}_n(t))\sqrt{g_{\mu\nu}\dot{x}_n^\mu\dot{x}_n^\nu}\,{\rm d}^4x}}\,.
\end{equation}

Notwithstanding this somewhat bizarre construction, the Lagrangian density maintains its (CP)T-oddness, essentially due to ${\rm d}s/{\rm d}t=\sqrt{g_{\mu\nu}\dot{x}^\mu\dot{x}^\nu}=1/u^0$. And this (CP)T-oddness is inherited by the so-called stress-energy tensor through its definition in terms of $\cal L$:
\begin{equation}
\label{eq.10}
T_{\mu\nu}=\frac{2}{\sqrt{-g}}\left[\frac{\partial(\sqrt{-g}\,{\cal L})}{\partial g^{\mu\nu}}-\frac{\partial}{\partial x^\lambda}\frac{\partial(\sqrt{-g}\,{\cal L})}{\partial{g^{\mu\nu}}_{,\lambda}}\right]\,,
\end{equation}
which indeed becomes
\begin{equation}
\label{eq.11}
T_{\rm S}^{\mu\nu}=\sum_n{m_n\frac{\delta^3({\bf x}-{\bf x}_n(t))}{\sqrt{-g}}\frac{u_n^\mu u_n^\nu}{u_n^0}}\,.
\end{equation}
The subscript `S' indicates that we are considering the field source, which is (C)PT-odd due to the presence of $u_n^0$, to distinguish it from the usual stress-energy tensor $T^{\mu\nu}$, which is (C)PT-even, and that can be obtained from $T_{\rm S}^{\mu\nu}$ by multiplying by a time integral over the Dirac delta function to get
\begin{equation}
\label{eq.12}
T^{\mu\nu}=\sum_n{m_n\int{\frac{\delta^4(x-x_n(s_n))}{\sqrt{-g}}u_n^\mu u_n^\nu\,{\rm d}s_n}}\,.
\end{equation}

Thus, while $T^{\mu\nu}$ does not change sign from matter to antimatter, $T_{\rm S}^{\mu\nu}$ does, as expected, since an antimatter-generated field must be opposite to a matter-generated one. In particular, the dominant component for a single non-relativistic particle is
\begin{equation}
\label{eq.13}
T_{\rm S}^{00}=m\frac{\delta^3({\bf x}-{\bf x}(t))}{\sqrt{-g}}u^0\,,
\end{equation}
i.e.\ essentially the gravitational charge of the particle \citep[see][]{vil11}, which is positive ($u^0={\rm d}t/{\rm d}s>0$) for matter, and negative ($u^0={\rm d}t/{\rm d}s<0$) for antimatter. On the contrary, the time-time component of the usual stress-energy tensor, i.e.\
\begin{equation}
\label{eq.14}
T^{00}=m\int{\frac{\delta^4(x-x(s))}{\sqrt{-g}}(u^0)^2\,{\rm d}s}\,,
\end{equation}
is positive definite, as it must be, representing the energy density.

If the Lagrangian density in Eq.~(\ref{eq.8}) includes all (matter and radiation) contributions to the field source, adding to it the (scalar) term for the space-time geometry, one gets the so-called Einstein-Hilbert action, 
\begin{equation}
\label{eq.15}
S=\int{\sqrt{-g}\left({\cal L}-\frac{R}{16\pi G}\right){\rm d}^4x}\,,
\end{equation}
which, through the action principle, yields the Einstein field equation
\begin{equation}
\label{eq.16}
R_{\mu\nu}-\frac{1}{2}g_{\mu\nu}R=8\pi GT^{\rm S}_{\mu\nu}\,.
\end{equation}

Let us consider an ensemble of matter and antimatter point particles where the radiation contribution to the source term $T^{\rm S}_{\mu\nu}$ is negligible and there is no contribution from a cosmological constant. From Eq.~(\ref{eq.11}), we can separate the matter and antimatter contributions as
\begin{eqnarray}
\label{eq.17}
T_{\rm S}^{\mu\nu}=T_{\rm SM}^{\mu\nu}+T_{\rm SA}^{\mu\nu}=\sum_M{m_M\frac{\delta^3({\bf x}-{\bf x}_M(t))}{\sqrt{-g}}\frac{u_M^\mu u_M^\nu}{u_M^0}}\nonumber\\+\sum_A{m_A\frac{\delta^3({\bf x}-{\bf x}_A(t))}{\sqrt{-g}}\frac{u_A^\mu u_A^\nu}{u_A^0}}=T_{\rm M}^{\mu\nu}-T_{\rm A}^{\mu\nu}\,.
\end{eqnarray}

Let us now suppose that the number and mass of the matter particles are (at least approximately) equal to those of the antimatter particles, so as to have equal total masses of matter and antimatter. If the differences in the $u^\mu$ components are also negligible, what keeps $T_{\rm S}^{\mu\nu}$ different from zero in Eq.~(\ref{eq.17}) (and then prevents the flat space-time of special relativity in Eq.~(\ref{eq.16})) is the spatial distribution of the various particles, represented by the Dirac delta functions. Only if we had all the matter and antimatter particles coupled two by two in the same space position, we would get $T_{\rm S}^{\mu\nu}=0$, since the gravitational field (or space-time curvature) produced by each matter particle would be canceled by the coupled antimatter particle or, in other words, there would be only neutral gravitational charges and no gravity at all. But this scenario is highly unlikely, since matter and antimatter repel each other (or, alternatively, they would annihilate each other).

Starting from a random distribution of particles, one can expect that, due to the matter and antimatter self-attraction and mutual repulsion, eventually matter and antimatter could be well separate in two distinct regions, possibly having first experienced a transient phase where more or less massive ``islands" of matter are distributed in space alternated with similar aggregations of antimatter. In this phase we would have matter ``particles" surrounded by antimatter ones, and vice versa, which means that each particle feels more the repulsive effect of the closer opposite particles than the attractive one from the more distant like particles, which gives rise to a global expansion and could also prevent the final aggregation into two distinct blocks. Anyway, we focus on this situation, which resembles that of our Universe, where matter is organized in superclusters separated by vast cosmic voids, in which, according to \citet{vil12b}, equivalent amounts of antimatter may be hidden\footnote{Previous studies on a matter-antimatter symmetric Universe \citep[e.g.][]{coh98} seemed to exclude the possibility that matter and antimatter domains have sizes smaller than the visible Universe, due to the lack of the expected annihilation radiation from the domain boundaries, but, unlike in those models, in our scenario annihilation is prevented by gravitational repulsion, so that such a lower limit is no longer valid.}.

In the standard general theory of relativity (i.e.\ that not extended to CPT-transformed matter), the field source in Eq.~(\ref{eq.16}) for an ideal fluid of matter point particles is usually expressed in terms of its energy density $\rho$ and isotropic pressure $p$ measured in the fluid rest frame, and of the fluid four-velocity $u^\mu$:
\begin{equation}
\label{eq.18}
T_{\rm Sfl}^{\mu\nu}=T_{\rm fl}^{\mu\nu}=(\rho+p)u^\mu u^\nu-g^{\mu\nu}p\,.
\end{equation}
This expression can be obtained by taking the average values over the particle population of the $T_{\rm S}^{\mu\nu}$ components in Eq.~(\ref{eq.11}). Thus, according to Eq.~(\ref{eq.17}), we have that $T_{\rm SAfl}^{\mu\nu}=-T_{\rm Afl}^{\mu\nu}$, i.e.\ also when $T_{\rm A}^{\mu\nu}$ is expressed as in Eq.~(\ref{eq.18}). As a consequence, it is clear that we can not adopt the simplifying version of the field source in terms of $\rho$ and $p$ of Eq.~(\ref{eq.18}) when dealing with two different kinds of field-generating particles, since in our above model of alternated matter and antimatter charges it would be null, totally ignoring the strong dependence on the particle distribution. We can guess that the appropriate source term for such a two-charge distribution is proportional to the expression given in Eq.~(\ref{eq.18}), which in our model is considered to be equal for matter and antimatter (in particular we have $\rho_{\rm M}=\rho_{\rm A}=\rho$), with $T_{\rm Mfl}^{\mu\nu}=T_{\rm Afl}^{\mu\nu}=T_{\rm SMfl}^{\mu\nu}=-T_{\rm SAfl}^{\mu\nu}$. Since we expect that the total effect is repulsive, we set
\begin{equation}
\label{eq.19}
T_{\rm Sfl}^{\mu\nu}=\alpha T_{\rm SAfl}^{\mu\nu}=-\alpha[(\rho+p)u^\mu u^\nu-g^{\mu\nu}p]\,,
\end{equation}
just to have $\alpha$ positive. The value of $\alpha$ will depend on the specific charge distribution, and we will explore possible values in the next section.

First we check that our guess in Eq.~(\ref{eq.19}) is correct in the ``generalized-Newtonian" limit (i.e.\ the Newtonian limit extended to the coexistence of positive and negative gravitational charges). Since in this limit the standard field equations (\ref{eq.16}) and (\ref{eq.18}) reduce to the standard Poisson's equation $\nabla^2\phi=4\pi G\rho$, our hypothesized source term in Eq.~(\ref{eq.19}) leads to a Poisson-like equation of the form
\begin{equation}
\label{eq.20}
\nabla^2\phi=-4\pi G\alpha\rho\,,
\end{equation}
which we must demonstrate to be appropriate.

The (generalized) Newtonian potential of a set of point particles with mass $m_n$ and gravitational charge $m_nu_n^0$ ($u_n^0=+1$ for matter and $-1$ for antimatter) felt at the space position $\bf x$ is
\begin{equation}
\label{eq.21}
\phi({\bf x})=-G\sum_n{\frac{m_nu_n^0}{|{\bf x}-{\bf x}_n|}}\,.
\end{equation}
Under the assumption that all particles have the same mass $m_n=m$, the Laplacian of Eq.~(\ref{eq.21}) at the location $\bf x$ of a given matter particle can be expressed by normalizing distances to the nearest neighbor one $|{\bf x}-{\bf x}_1|$ as
\begin{equation}
\label{eq.22}
\nabla^2\phi({\bf x})=G\alpha m\nabla^2\frac{1}{|{\bf x}-{\bf x}_1|}=-4\pi G\alpha m\delta^3({\bf x}-{\bf x}_1)\,,
\end{equation}
where $\alpha$ is the dimensionless constant
\begin{equation}
\label{eq.23}
\alpha=-\sum_n{\frac{u_n^0|{\bf x}-{\bf x}_1|}{|{\bf x}-{\bf x}_n|}}\,.
\end{equation}
In a periodic lattice distribution, each elemental cell of size $\sim2|{\bf x}-{\bf x}_1|$ contains a mass $m$ of matter and a mass $m$ of antimatter, so that $m\delta^3({\bf x}-{\bf x}_1)$ in Eq.~(\ref{eq.22}) represents the density $\rho$ of one of the two components, and Eq.~(\ref{eq.22}) reduces to Eq.~(\ref{eq.20}), QED.

\section{Cosmological equations}

The electrostatic counterpart of the dimensionless constant $\alpha$ is the so-called Madelung constant \citep[after the work of][]{mad18}, used for the calculation of the binding energy of ionic crystal lattices\footnote{This simile between ionic crystals and Universe structure can also be found in \citet{rip10}.}. Since the infinite summation of our ``gravitational" $\alpha$ is formally identical to that of the Madelung constant, we can refer to the already computed values for specific crystal structures. For example, the value for the sodium-chloride (rock-salt) structure (i.e.\ that originally calculated by Madelung), which is composed of two interpenetrating face-centered cubic lattices, one for each of the two ion types, ${\rm Na}^+$ and ${\rm Cl}^-$, is now known to thousands of decimal digits: $|\alpha_{\rm NaCl}|\approx1.74756$. Another simple and highly symmetric structure is that of the cesium chloride (two interpenetrating cubic lattices resulting in a kind of body-centered cubic structure, with each ion at a cube center surrounded by eight opposite ions at the vertices of the cube). In this latter case the value of the Madelung constant is $|\alpha_{\rm CsCl}|\approx1.76267$. Two crystal variants of ZnS (zinc blende or sphalerite, and wurtzite) give both $|\alpha_{\rm ZnS}|\approx1.64$, with a slight difference. When considering also more complex crystal structures, $|\alpha|$ is typically found to vary between 1.5 and 2.5, which we can take as our ``trial" range.

With the Friedmann-Lema\^itre-Robertson-Walker metric for a spatially homogeneous and isotropic expanding Universe with scale factor $a(t)$, the time-time component of the field equation (\ref{eq.16}) for a matter-antimatter lattice Universe (i.e.\ with the source term given by Eq.~(\ref{eq.19})) becomes
\begin{equation}
\label{eq.24}
\frac{\ddot{a}}{a}=\frac{4}{3}\pi G\alpha(\rho+3p)\,,
\end{equation}
which clearly shows that in our model, where $\alpha$ was set to be positive (as confirmed by Eq.~(\ref{eq.23})), the Universe expansion is accelerated ($\ddot{a}>0$). In the dust approximation (i.e.\ no relativistic matter or antimatter), the pressure $p$ vanishes, and Eq.~(\ref{eq.24}) can be interpreted in generalized-Newtonian terms as the (mean) acceleration felt at the surface of a sphere of radius $a$ due to the resulting gravitational effect of the matter-antimatter content of the sphere. By adding the two opposite contributions from matter and antimatter as they would be individually (whose sum is equal to zero), Eq.~(\ref{eq.24}) in the dust approximation can be rewritten as
\begin{equation}
\label{eq.25}
\ddot{a}=(1+\alpha)\frac{GM_{\rm A}}{a^2}-\frac{GM_{\rm M}}{a^2}\,,
\end{equation}
where $M_{\rm M}=M_{\rm A}=4\pi a^3\rho/3$ is the mass/energy of matter (and of antimatter) contained in the sphere. Thus, we can see that the repulsive contribution is $(1+\alpha)$ times the attractive one. With $\alpha$ in our fiducial range 1.5--2.5, this ratio is similar to that found in the $\Lambda$CDM model between dark energy and matter.

A second cosmological equation can be obtained by subtracting Eq.~(\ref{eq.24}) from the space-space component of the field equation (but can also be derived in other ways), to eliminate both $\ddot{a}$ and $p$:
\begin{equation}
\label{eq.26}
\left(\frac{\dot{a}}{a}\right)^2=-\frac{8}{3}\pi G\alpha\rho-\frac{k}{a^2}\,,
\end{equation}
with $k<0$, $=0$, $>0$ for negative, zero, positive spatial curvature, respectively. It is evident that in our case $k<0$, in contrast with the current standard model for a spatially flat Universe\footnote{Although observations seem to favor $k=0$, a negatively curved space can not be excluded \citep[see e.g.][]{ben12}.}. By rearranging the terms in Eq.~(\ref{eq.26}), one gets
\begin{equation}
\label{eq.27}
\frac{\dot{a}^2}{2}+\frac{4}{3}\pi G\alpha\rho\frac{a^3}{a}=\frac{|k|}{2}\,,
\end{equation}
which has the form of the conservation law of the total (kinetic plus potential) mechanical energy per unit mass, equal to $|k|/2$. Using the same trick as in Eq.~(\ref{eq.25}), we can see that also here the ratio between the repulsive potential energy and the attractive one is equal to $1+\alpha$. From the $\nu=0$ component of the stress-energy conservation law $T^\mu_{\nu;\mu}=0$, in the dust approximation one gets the energy conservation equation $\rho a^3=\rm constant$. Thus, the potential in Eq.~(\ref{eq.27}) tends to zero as $a\rightarrow\infty$, and $\dot{a}\rightarrow\sqrt{|k|}$. In this dust-dominated phase there is no initial singularity, since $\dot{a}=0$ at a certain initial time $t_{\rm i}$ when $a_{\rm i}\equiv a(t_{\rm i})=8\pi G\alpha\rho_0/3|k|$ (where as usual `0' subscripts indicate the present values and we have fixed $a_0=1$). Even though the critical density $\rho_{\rm cr}=3H^2/8\pi G$ (with $H=\dot{a}/a$) has no physical meaning in our model, it nevertheless represents a useful and familiar normalization for the density parameter $\Omega=\rho/\rho_{\rm cr}$. Thus, in terms of $H_0$ and $\Omega_0$, Eq.~(\ref{eq.26}) becomes
\begin{equation}
\label{eq.28}
\dot{a}^2=H_0^2(1+\alpha\Omega_0)\left(1-\frac{a_{\rm i}}{a}\right)\,,\quad a_{\rm i}=\frac{\alpha\Omega_0}{1+\alpha\Omega_0}\,.
\end{equation}

\section{Age of the Universe and SN Ia test}

\begin{figure*}
\centering
\includegraphics[width=13cm]{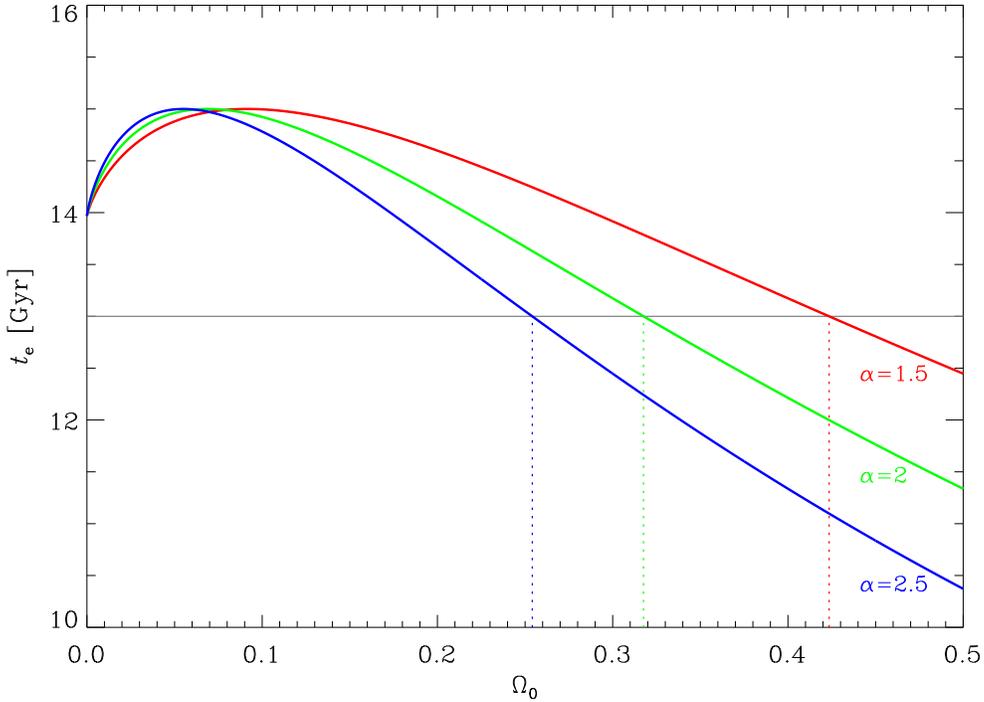}
\caption{The expansion age of the dust-dominated matter-antimatter Universe as a function of the density parameter $\Omega_0$ for various values of $\alpha$: 1.5 (red), 2 (green), 2.5 (blue); the gray line indicates a possible lower limit of $13\rm\,Gyr$.}
\label{fig:1}
\end{figure*}

The ages of the oldest stars in globular clusters constrain the age of the Universe in the range 12--$15\rm\,Gyr$ \citep[e.g.][]{kra03}, while cosmic microwave background (CMB) anisotropy and large-scale structure measurements give a model-dependent age of $13.8\pm0.2\rm\,Gyr$ for a flat Universe \citep[e.g.][]{teg06,jar11,kom11,sul11}.

In our dust-dominated, matter-antimatter model, we do not find any singularity, so that we can speak of ``age of the expansion" rather than of ``age of the Universe", not knowing when a possible ``birth" happened with respect to the start of the dust-dominated expansion. However, we may suppose that the expansion age is close to what must be compared with observational results. From Eq.~(\ref{eq.28}), the expansion age, i.e.\ the time elapsed since the start of the model expansion, is
\begin{equation}
\label{eq.29}
t_{\rm e}\equiv t_0-t_{\rm i}=\frac{1}{H_0\sqrt{1+\alpha\Omega_0}}\int_{a_{\rm i}}^1{\frac{{\rm d}a}{\sqrt{1-a_{\rm i}/a}}}\,.
\end{equation}

The result of the integration in Eq.~(\ref{eq.29}) is a complicated analytic function of $\alpha\Omega_0$. With $H_0=70\rm\,km\,s^{-1}\,Mpc^{-1}$, in Fig.~\ref{fig:1} we plot the expansion age $t_{\rm e}$ as a function of $\Omega_0$ for three values of $\alpha$ in the confidence interval: 1.5, 2, 2.5. The function starts from $1/H_0\approx13.97\rm\,Gyr$ at $\Omega_0=0$ and reaches a maximum of $t_{\rm e,max}\approx15.00\rm\,Gyr$ at $\Omega_0\approx0.092$, 0.069, and 0.055 for $\alpha=1.5$, 2, and 2.5, respectively. Then it decreases with increasing $\Omega_0$. As one can see from the figure, the expansion age does not yield severe constraints on $\Omega_0$. Indeed, by setting a lower limit to $t_{\rm e}$ of, e.g., $13\rm\,Gyr$ (gray line), the tightest constraint is achieved for $\alpha=2.5$ as $\Omega_0\lesssim0.25$, i.e.\ nothing particularly unexpected, apart from the fact that any lower value (even a baryonic-only Universe with $\Omega_0\lesssim0.05$) is not excluded by the oldest-stars constraint.

\begin{figure*}
\centering
\includegraphics[width=13cm]{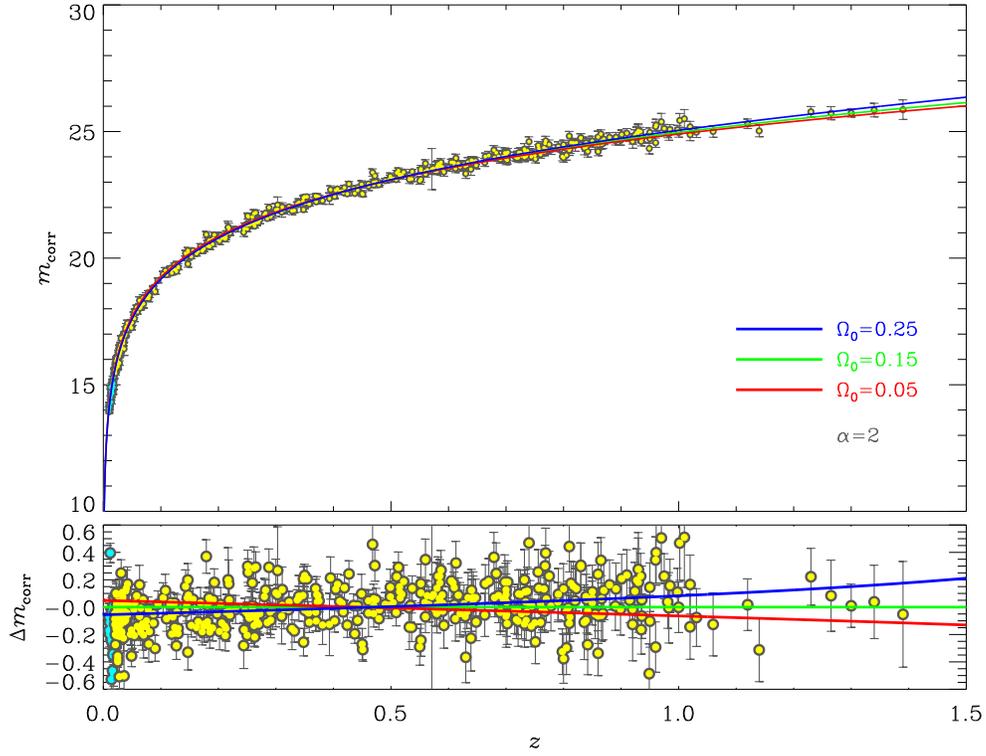}
\caption{Hubble diagram of SNe Ia from the data set of \citet{con11}; cyan symbols refer to $z<0.02$ events, which are excluded from the fitting procedure. Matter-antimatter model fits with $\alpha=2$ and $\Omega_0=0.05$ (red), 0.15 (green), 0.25 (blue) are overplotted; residuals are shown in the bottom panel.}
\label{fig:2}
\end{figure*}

\begin{figure*}
\centering
\includegraphics[width=13cm]{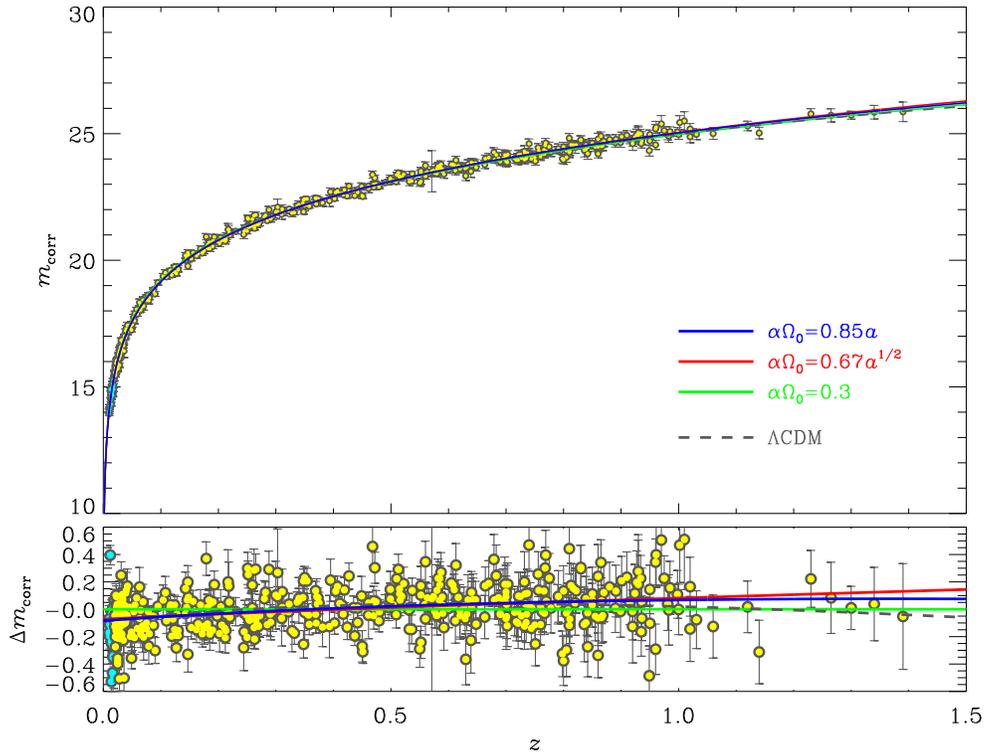}
\caption{The same as in Fig.~\ref{fig:2}, but with model fits with $\alpha$ variable as $\alpha\propto\sqrt{a}$ (red) and $\alpha\propto a$ (blue), compared with an $\alpha=\rm constant$ model from Fig.~\ref{fig:2} (green) and the $\Lambda$CDM model (dashed).}
\label{fig:3}
\end{figure*}

Another very important cosmological test is the Hubble diagram of Type Ia supernovae (SNe Ia), which is so well fitted by the $\Lambda$CDM model with $\Omega_{\rm M}$ about 0.27 ($\Omega_{\Lambda}\sim0.73$), being the historical proof of the expansion acceleration. Indeed, since the milestone works of \citet{rie98} and \citet{per99}, it was evident that distant SNe Ia are fainter than expected from a decelerating Universe.

In Fig.~\ref{fig:2} we plot the corrected apparent magnitudes of the 472 SNe of the data set from \citet{con11} versus their redshift $z$. For the magnitude correction, $m_{\rm corr}=m_B+\alpha_{\rm n}(s-1)-\beta_{\rm n}{\cal C}$, we adopted the same nuisance parameters found by \citet{con11} and \citet{sul11} in their $\chi^2$ minimization fits: $\alpha_{\rm n}=1.37$, $\beta_{\rm n}=3.18$. From Eq.~(\ref{eq.28}) we derive the luminosity distance in our model as
\begin{eqnarray}
\label{eq.30}
d_L(z)=\frac{1+z}{H_0\sqrt{1+\alpha\Omega_0}}\int_0^z{\frac{{\rm d}z'}{(1+z')\sqrt{1-a_{\rm i}(1+z')}}}\nonumber\\=\frac{2(1+z)}{H_0\sqrt{1+\alpha\Omega_0}}\left[\tanh^{-1}\sqrt{1-a_{\rm i}(1+z')}\right]_z^0\,,
\end{eqnarray}
which, with $H_0=70\rm\,km\,s^{-1}\,Mpc^{-1}$ and $\alpha=2$, gives the three plotted fits for $\Omega_0=0.05$, 0.15, and 0.25. Due to the various well-known problems affecting low-redshift SNe Ia \citep[see e.g.][]{rie07,kes09,con11}, we excluded events with $z<0.02$ (36 out of 472, cyan symbols in the figure) from the fitting procedure. The residuals in the bottom panel show comparably good fits, even if $\Omega_0=0.25$ yields a better $\chi^2_{\rm red}$ (1.55) due to the closer agreement with the large number of low-$z$ SNe. (An even lower $\chi^2_{\rm red}=1.53$ is obtained for $\Omega_0=0.28$, not shown in the figure.) Although we excluded the closest SNe, the remaining nearby SNe (say, $z\lesssim0.1$) could still be affected by systematic uncertainties \citep[see e.g.][]{kel10,ben12}, so that we can not rule out low values of $\Omega_0$, which actually fit better the most distant ($z\gtrsim1$) events, and even a baryonic-only Universe can not be excluded.

On the contrary, if low-$z$ SNe Ia should not be affected by significant bias, it seems that the model fits should be more curved (especially at low $z$) to better match the data. In this regard, we recall that our simple model has a main limitation, since it describes only the dust-dominated phase in the Universe history, and this phase would stop in the past when $a=a_{\rm i}$. Even considering the most favorable case $\Omega_0=0.05$ and $\alpha=1.5$, $a_{\rm i}$ would not be smaller than $\sim0.07$, i.e.\ by far too large to meet the epoch producing the CMB radiation. We can guess that $\alpha$ must decrease with decreasing time while matter becomes hotter, and this would allow the decrease of $a_{\rm i}$ too towards arbitrarily small values. Just for exercise and without any serious intention of modeling, we can see what would happen with a toy model where $\alpha$ changes with time in Eq.~(\ref{eq.28}) as $\alpha=\alpha_0a^\beta$.

In the Hubble diagram of Fig.~\ref{fig:3} we plot the $\Lambda$CDM model fit with $\Omega_{\rm M}=0.27$ compared with three matter-antimatter models. The $\alpha=\rm constant$ model ($\alpha\Omega_0=0.3$) is the same of Fig.~\ref{fig:2} with $\alpha=2$ and $\Omega_0=0.15$ and is taken as a reference for an easier comparison between the bottom panels of the two figures. We see that with $\alpha=\alpha_0a^\beta$ ($\beta=1/2$, 1 in our examples) we obtain the above-mentioned curvature to better fit the low-$z$ data, similarly to what happens with the $\Lambda$CDM model, which indeed fits the data with a $\chi^2_{\rm red}$ (1.51) lower than those found for the $\alpha=\rm constant$ models (1.53--1.55). Comparable (or better) fits with a variable $\alpha$ are obtained with $\alpha_0\Omega_0=0.67$ for $\alpha=\alpha_0\sqrt{a}$ ($\chi^2_{\rm red}=1.51$) and with $\alpha_0\Omega_0=0.85$ for $\alpha=\alpha_0a$ ($\chi^2_{\rm red}=1.49$). With, e.g., $\Omega_0=0.25$, in the former case $\alpha$ would vary between 2.68 and 1.69 from $z=0$ to $z=1.5$, while in the latter the variation range would be $\alpha=3.40$--1.36, which appears too wide. In any case, as previously found in the $\alpha=\rm constant$ models, also here high values of $\Omega_0$ give lower $\chi^2_{\rm red}$'s. However, the closeness of these $\chi^2_{\rm red}$ values among the various models (including the $\Lambda$CDM model) and, even more, the existing issue on the low-$z$ data, do not allow us to discriminate among them on this basis. We notice that all matter-antimatter model fits diverge from the $\Lambda$CDM one towards high $z$, predicting significantly fainter SN apparent magnitudes at $z\gtrsim1.5$. Thus, new data at larger distances, or possible revised data for nearby events, should eventually allow us to disentangle.

\section{Discussion and conclusions}

In the previous section we have submitted our matter-antimatter cosmological model to two specific tests: the Universe age and the SN Ia test, both passed successfully. There are other observational constraints that should be checked to be in agreement with the model, such as the primordial abundances of light elements and the acoustic scale of the CMB. Both these constraints have been discussed and investigated in detail by \citet{ben12} in their ``Dirac-Milne" cosmology, i.e.\ a matter-antimatter model that appears as a limit case of ours: the one with $\alpha=0$. Since we have already noticed that in our cosmological model $\alpha$ is expected to become very small or null in the early stages of the Universe pertinent to primordial nucleosynthesis and CMB, we can rely on those results, and possibly postpone a detailed study of these issues to future works.

Regarding the radiation-dominated era preceding the matter-antimatter dust-dominated one, i.e.\ when pressure is no longer negligible and the energy conservation law becomes $\rho a^4=\rm constant$, one can easily check from Eq.~(\ref{eq.26}) that also in this case there is no initial singularity, unless $\alpha\rightarrow0$ with a certain rapidity as $t\rightarrow t_{\rm i}$. Thus, we can conclude that in our model no singularity is required, but arbitrarily small initial Universe sizes are allowed. Another important feature of this model is the absence of the horizon problem, since the scale factor acceleration has never been negative, and in the earliest stages $\dot{a}$ can even approach zero.

Unfortunately, the various tests can not provide strong constraints on the value of $\Omega_0$, but all values between $\sim0.05$ (baryonic-only Universe) and $\sim0.25$ (existence of dark matter) are possible, even though the higher values seem to be favored by lower $\chi^2_{\rm red}$'s in the SN Ia Hubble diagram fits. In any case, in our repulsive-gravity scenario, there seems to be no need for mysterious matter in addition to the well-known baryonic matter to explain the phenomena for which dark matter is usually invoked. Indeed, as shown by \citet{haj11c,haj12a}, the presence of additional unseen matter at galactic scales that would explain, e.g., the galaxy rotational curves, can be successfully replaced by the effect of the gravitational polarization of the quantum vacuum induced by the galaxy mass. On larger, cluster scales, the ``observation" of potential wells deeper than expected from baryonic matter alone, which would allow clustering and would produce the weak gravitational lensing \citep[see e.g.\ the recent observations in][]{die12}, could just be the effect of the presence of surrounding potential ``hills" due to antimatter in the adjacent voids.

In summary, starting from the basic assumptions that antimatter is CPT-transformed matter and that our Universe is matter-antimatter symmetric, we have developed a cosmological model where, consistently with general relativity, gravitational repulsion between the two opposite components is the cause of the accelerated expansion of the Universe. This has been done neither with modifications to existing well-established theories, nor with the ad-hoc introduction of unknown entities and dark ingredients. Due to the evident absence of matter in the well-known cosmic voids, these are the favorite candidates to host antimatter, whose invisibility has been discussed and motivated in previous works. The resulting lattice structure is well reported in the current three-dimensional maps of the observed Universe. While in an electrostatic lattice structure (i.e.\ a crystal) the alternation of opposite charges (whose interaction is attractive) provides a net binding energy in spite of a null total charge, the alternance of unlike gravitational charges in the cosmos produces a net accelerated expansion in spite of the equal amounts of the two components. Similarly to the Madelung constant in crystals, the degree of resulting repulsive energy is measured by the parameter $\alpha$, which we supposed to be in the range 1.5--2.5, and which multiplies the matter (or antimatter) energy density in the cosmological equations, thus together providing a single parameter. The ratio between the repulsive and attractive energies is equal to $1+\alpha$, i.e.\ very close to that found between dark energy and matter in the $\Lambda$CDM model, thus solving the coincidence problem mentioned in the Introduction. In contrast to the standard model, the acceleration has never been negative and horizon and singularity problems are absent in our model. With $\alpha$ in the above confidence interval and with $\Omega_0$ in the range 0.05--0.25, the age of the Universe varies between about 13 and $15\rm\,Gyr$. Model fits to the SN Ia Hubble diagram are comparable with that of the $\Lambda$CDM model in the observed range $z<1.4$, while they diverge (SNe fainter for our model) at higher redshifts, thus offering a future test to discriminate between them. Besides dark energy, even the existence of dark matter is not needed in our scenario, though it is allowed, maybe favored by the SN Ia test.

The standard $\Lambda$CDM model is currently the simplest and most popular attempt to explain the cosmic acceleration, identifying dark energy with the cosmological constant. There exists a wide variety of alternative, competing models, which are usually more sophisticated but equally compatible with observational constraints. They invoke ``dark fluids", scalar fields, or geometrical modifications to general relativity \citep[see e.g.\ the recent review by][]{bam12}. However, all of them have the same problem of the simpler $\Lambda$CDM model, i.e.\ no physical justification for the new unknown ingredients or geometries, beyond the consistency of the models with observational data.

In conclusion, from the theoretical point of view our model appears more elegant and self-consistent than the current dark-energy cosmologies, being based only on well-known physical entities and theories, with no need for ad-hoc, unknown but dominant, components. Moreover, it spontaneously solves several heavy issues like the horizon and coincidence problems, the initial singularity, the apparent matter-antimatter asymmetry. On the other hand, at the moment we lack experimental confirmation for the predicted repulsive gravity between matter and antimatter, but we hope in an answer in a few years from the ongoing experiments \citep[e.g.][]{kel08}.

\makeatletter
\let\clear@thebibliography@page=\relax
\makeatother



\end{document}